\documentclass[prx,twocolumn,preprintnumbers,amsmath,
superscriptaddress,amssymb]{revtex4}

\usepackage{graphicx}% Include figure files
\usepackage{bm}% bold math
\usepackage{color}% bold math
\usepackage{ulem}

\begin{document}

%\preprint{APS/123-QED}

\title{Ultrafast photoluminescence in metals: Theory and its application to silver}
\author{Shota Ono}
\email{shota\_o@gifu-u.ac.jp}
\affiliation{Department of Electrical, Electronic and Computer Engineering, Gifu University, Gifu 501-1193, Japan}
\author{Tohru Suemoto}
\affiliation{Toyota Physical and Chemical Research Institute, Yokomichi 41-1, Nagakute-shi Aichi, 4801192, Japan}

\begin{abstract}
We study the transient photoluminescence (PL) of photoexcited metals by solving the Boltzmann equation considering the effects of electron-electron (e-e) and electron-phonon (e-ph) collisions, where the e-ph coupling function is calculated from first-principles in order to account for the energy transfer rate between electrons and phonons accurately. We apply the present scheme to the transient PL of silver and demonstrate that the agreement between the theory and experiment is good, where the effect of nonequilibrium electron distribution is significant to fit the experimental data. The effects of the nanoscale roughness at metal surfaces and the e-e umklapp scattering on ultrafast electron dynamics are also discussed. 
\end{abstract}

%\pacs{}

\maketitle

%%%%%%%%%%%%%%%%%%%%%%%%%%%%%%%%%%%%
\section{Introduction}
The ultrafast electron and phonon dynamics in photoexcited solids has been extensively studied since the development of ultrafast laser pulse. After the absorption of laser pulse, excited electrons interact with each other as well as lattice vibrations and will obey the quasiequilibrium distribution that is characterized by an electron temperature much higher than the lattice temperature. One of the important questions is, beyond the quasiequilibrium approximation used in the two-temperature model \cite{allen}, to understand how nonequilibrium electron and phonon distributions influence the transient optical properties of solids. For example, the Boltzmann equation \cite{mueller,kabanov2014,waldecker,ono2017,pablo2017,rethfeld,ono2018,pablo2020} must be solved to study the time-evolution of nonequilibrium distributions in excited solids, which is applied to interpret pump-probe experiments. 

%It is usually assumed that the effect of spatial diffusion is absent or negligibly small compared to that of scattering or collision. This is true when the electron dynamics in thin films are considered because no diffusion occurs perpendicular to the film surface. This may also be true when the electron-phonon (e-ph) coupling is strong enough to allow the excess energy to be transferred rapidly to lattice vibrations or phonons near the laser-irradiated surface. However, this is questionable otherwise. In practice, the diffusion equation approach has been used to study ultrafast dynamics \cite{juhasz,maris,oliver}, while the electron and lattice temperatures are assumed to be present at any time. The material-dependent parameters are reduced to macroscopic quantities only; the specific heat and the e-ph coupling constant. 

Recently one of the authors has reported femtosecond infrared photoluminescence (PL) in a noble metal, gold \cite{suemoto2019}. The time-evolution of the PL spectra has been successfully reproduced by an effective temperature model also accounting for nonequilibrium component (i.e., the deviation from thermal equilibrium) in a phenomenological way. In the present work, we apply the standard Boltzmann equation approach to the PL decay of silver, where the electron band structure of silver is more simple than that of gold because the location of $d$ band is deeper in the former. The electron and phonon band structures and the electron-phonon (e-ph) coupling function (i.e., the Eliashberg function \cite{grimvall}) are calculated from first-principles. We demonstrate that the calculated PL curves are in good agreement with the experiment. The condition of electron nonequilibrium is important to explain the experimental data. By tuning the free parameters of excitation density and effective electron-electron (e-e) interaction strength, we discuss the effects of the nanoscale roughness at metal surfaces and the e-e umklapp scattering on the ultrafast electron dynamics.

%Some deviations may be attributed to the electron 

Below, the Boltzmann equation approach with the use of {\it ab initio} calculations are described in Sec.~\ref{sec:theory}; Numerical results of the PL decay dynamics for several excitation parameters are provided in Sec.~\ref{sec:R1}; Application to experimental results is provided in Sec.~\ref{sec:R2}; and the conclusion is given in Sec.~\ref{sec:conclusion}.

%%%%%%%%%%%%%%%%%%%%%%%%%%%%%%%%%%%%
\section{Theory}
\label{sec:theory}
%%%%%%%%%%%%%%%%%%%%%%%%%%%%%%%%%%%%
Following Ref.~\cite{ono2018}, we study the time-evolution of the electron distribution $f(\varepsilon)$ and the phonon distribution $n(\omega)$, where $\varepsilon$ and $\omega$ are the electron energy and the phonon frequency, respectively. Such distributions can be obtained by taking the wavevector average of distribution functions. The time-evolution of $f(\varepsilon)$ and $n(\omega)$ are calculated by solving the Boltzmann equation
\begin{eqnarray}
\frac{\partial f (\varepsilon)}{\partial t}
&=& \left( \frac{\partial f }{\partial t}\right)_{\rm e-e} 
+ \left( \frac{\partial f }{\partial t}\right)_{\rm e-ph}
+ \left( \frac{\partial f }{\partial t}\right)_{\rm laser},
\label{eq:Boltzmann_el}
\\
\frac{\partial n (\omega)}{\partial t}
&=& \left( \frac{\partial n }{\partial t}\right)_{\rm ph-e},
%+ \left( \frac{\partial n }{\partial t}\right)_{\rm ph-ph}
\label{eq:Boltzmann_ph}
\end{eqnarray}
where the right hand sides denote the collision integrals for the e-e, e-ph, and phonon-electron (ph-e) scattering. The contribution from the phonon-phonon scattering can be ignored when we focus on the electron dynamics within a few ps. The laser excitation term is also introduced and will be treated phenomenologically. 

%The electron diffusion term will be introduced in Sec.~\ref{sec:R2}.

As explained in Ref.~\cite{beversluis}, the photon absorption and emission processes must occur with the help of the localized surface plasmons or defects. To model such an intraband transition, we may consider the energy conservation law only: The wavevector conservation law is not taken into account explicitly in the laser excitation (Sec.~\ref{sec:laser}) and the PL intensity (Sec.~\ref{sec:PL}) calculations. We ignore the phonon contribution to the absorption and emission processes because the phonon energy is much smaller than the photon energy in the present situation.

%%%%%%%%%%%%%%%%%%%%%%%%%%%%%%%%%%%%
\subsection{Electron-electron collision}
The nonequilibrium electron distribution is created by the absorption of light (see Sec.~\ref{sec:laser}) and is redistributed through e-e collision events, yielding the electron quasiequilibrium state characterized by the time-dependent electronic temperature. The collision process is denoted by $(\varepsilon + \varepsilon') \rightarrow (\xi + \xi')$, where $\varepsilon $ and $\varepsilon'$ ($\xi$ and $\xi'$) are the single-particle electron energies of the initial (final) state. The e-e collision term for $f(\varepsilon)$ is thus given by three-dimensional integral for $\varepsilon', \xi$, and $\xi'$:
\begin{widetext}
\begin{eqnarray}
\left( \frac{\partial f }{\partial t}\right)_{\rm e-e}
&=& 2\pi\int d\varepsilon' \int d\xi \int d\xi'  
C_{\rm e-e}(\varepsilon,\varepsilon',\xi,\xi')
\delta (\varepsilon + \varepsilon' - \xi - \xi')
\nonumber\\
&\times&
\left\{ 
- f (\varepsilon) f (\varepsilon' ) [1- f(\xi)] [1- f(\xi' )]
+ [1- f (\varepsilon)] [1- f (\varepsilon' )] f(\xi) f(\xi' )
\right\},
\end{eqnarray}
where the coupling function is written as \cite{kabanov2014,ono2018}
\begin{eqnarray}
C_{\rm e-e}(\varepsilon,\varepsilon',\xi,\xi')
&=& \frac{1}{\hbar N(\varepsilon)}
\sum_{\bm{k}_1,\bm{k}_2,\bm{k}_{3},\bm{k}_4}\vert v_{\rm sc}(\bm{k}_3 - \bm{k}_1) \vert^2
\nonumber\\
&\times&
\delta (\varepsilon - \varepsilon_{\bm{k}_1})
\delta (\varepsilon' - \varepsilon_{\bm{k}_2})
\delta (\xi - \varepsilon_{\bm{k}_3})
\delta (\xi' - \varepsilon_{\bm{k}_4})
\delta_{ \bm{k}_1+\bm{k}_2,\bm{k}_{3}+\bm{k}_4},
\label{eq:cee}
\end{eqnarray}
where $\hbar$ is the Planck constant, $N(\varepsilon)$ is the electron density-of-states (DOS) per spin, $\bm{k}_i \ (i=1,\cdots,4)$ is the electron wavevector, and $v_{\rm sc}$ is the Fourier transform of the screened Coulomb interaction potential. Assuming that the Coulomb scattering rate is isotropic, i.e., independent of the wavevector, and using the expression of $N(\varepsilon)= \sum_{\bm{k}} \delta (\varepsilon - \varepsilon_{\bm{k}})$, one obtains
\begin{eqnarray}
\left( \frac{\partial f }{\partial t}\right)_{\rm e-e}
&=& \frac{2\pi}{\hbar} v_{0}^{2} \int d\varepsilon' \int d\xi \int d\xi'  
\delta (\varepsilon + \varepsilon' - \xi - \xi')
N(\varepsilon') N(\xi) N(\xi')
\nonumber\\
&\times&
\left\{ 
- f (\varepsilon) f (\varepsilon' ) [1- f(\xi)] [1- f(\xi' )]
+ [1- f (\varepsilon)] [1- f (\varepsilon' )] f(\xi) f(\xi' )
\right\},
\label{eq:ee2}
\end{eqnarray}
%\end{widetext}
where $v_0$ is the effective Coulomb interaction strength derived from Eq.~(\ref{eq:cee}). In the present study, the magnitude of $v_0$ is the first parameter to be determined from experiment. The expression of Eq.~(\ref{eq:ee2}) is fundamentally the same as that derived in Ref.~\cite{wais}. 

%quite time consuming. In the present study, the relaxation time approximation is used for the e-e collision term. This is given by
%\begin{eqnarray}
%\left( \frac{\partial f }{\partial t}\right)_{\rm e-e} 
%= - \frac{f(\varepsilon) - f_{\rm FD}(T_{\rm e})}{\tau_{\rm e-e}},
%\label{eq:ee}
%\end{eqnarray}
%where $\tau_{\rm e-e}$ is the relaxation time toward the Fermi-Dirac (FD) distribution function $f_{\rm FD}(T_{\rm e})$ with $T_{\rm e}$. The chemical potential is determined by the electron number conservation. In the following, $\tau_{\rm e-e}$ is treated as a numerical parameter.

%%%%%%%%%%%%%%%%%%%%%%%%%%%%%%%%%%%%
\subsection{Electron-phonon and phonon-electron collision}
\label{sec:eph}
The excess electron energy decreases and increases by emitting and absorbing phonons, respectively. The electron and phonon distributions are then redistributed simultaneously with the total energy conserved. These processes are described by the e-ph and ph-e collision terms
%\begin{widetext}
\begin{eqnarray}
%%%%%%%%%%%%%%%%
\left( \frac{\partial f }{\partial t}\right)_{\rm e-ph}
&=& 2\pi \frac{N(\varepsilon_F)}{N(\varepsilon)} \int d\xi \int d\omega  \alpha^2 F (\omega) 
\left[
\delta (\varepsilon  - \xi - \hbar\omega) S_1(\varepsilon, \xi, \omega)
+ 
\delta (\varepsilon  - \xi + \hbar\omega) S_2(\varepsilon, \xi, \omega)
\right],
\\
%%%%%%%%%%%%%%%%
\left( \frac{\partial n }{\partial t}\right)_{\rm ph-e} 
&=& 4\pi \frac{N(\varepsilon_F)}{D (\omega)}\int d\varepsilon \int d\xi  \alpha^2F(\omega)
f (\varepsilon) [1- f(\xi)]
\left\{
- n(\omega) \delta (\varepsilon - \xi + \hbar \omega)
+ [n(\omega)+1] \delta (\varepsilon - \xi - \hbar \omega)
\right\},
%%%%%%%%%%%%%%%%
\end{eqnarray}
\end{widetext}
where $\alpha^2F(\omega)$ is the Eliashberg function and $D(\omega)$ is the phonon DOS. The factor $N(\varepsilon_F)/N(\varepsilon)$ is a correction term of $\alpha^2F(\omega)$ for the electron energy deviation from the Fermi energy $\varepsilon_F$ \cite{ono2018}. The $f$ and $n$ enter into $S_1$ and $S_2$ given by
\begin{eqnarray}
S_1 &=& [f(\xi) - f (\varepsilon)] n(\omega) - f (\varepsilon) [1- f(\xi )],
\\
S_2 &=& [f(\xi) - f (\varepsilon)] n(\omega) + f(\xi ) [1- f (\varepsilon)].
\end{eqnarray}
The material-dependent functions, that is, $N(\varepsilon)$, $D(\omega)$, and $\alpha^2F(\omega)$, are obtained from {\it ab initio} calculations described in Sec.~\ref{sec:comp}.

%It should be noted that the e-ph coupling is weak in noble metals in general. For example, we consider the second moment of the Eliashberg function that is a measure of the e-ph energy transfer rate \cite{allen}
%\begin{eqnarray}
% \lambda \langle \omega^2 \rangle = 2 \int \alpha^2F(\omega)\omega d\omega.
%\end{eqnarray}
%According to Ref.~\cite{lin}, $\lambda\langle \omega^2 \rangle = 22.5$ meV$^2$ for silver and $\lambda\langle \omega^2 \rangle=185.9$ meV$^2$ for aluminum, while these are extracted from experimental data not from {\it ab initio} calculations. As a consequence of the weak e-ph coupling of silver, we will introduce another cooling mechanism in order to explain the PL spectra observed experimentally in Sec.~\ref{sec:R2}. 

%%%%%%%%%%%%%%%%%%%%%%%%%%%%%%%%%%%%
%\subsection{phonon-phonon collision}
%The phonon distribution function is disturbed by an energy transfer through the e-ph scattering. The nonequilibrium distribution can relax toward Bose-Einstein (BE) function by ph-ph collision events. This effect is important in studying relatively slow dynamics (the order of hundred picoseconds). We will show that the ph-ph collision effect can be negligible for investigating the PL dynamics within a few ps, by assuming an extreme situation, where the phonon distribution function keeps the BE function with $T=0$ K at each $t$

%%%%%%%%%%%%%%%%%%%%%%%%%%%%%%%%%%%%
\subsection{Electron-photon collision}
\label{sec:laser}
The laser pulse absorption is modeled by adding the collision integral
\begin{eqnarray}
\left( \frac{\partial f}{\partial t}\right)_{\rm laser}
  &=& P_0 G(t,t_p) Q(\varepsilon,\Omega_{\rm in}),
  \label{eq:laser1}
\end{eqnarray}
where $P_0 \ (\ge 0)$ is a parameter for the intensity of the gaussian-type pulse peaked at $t_p$ and width $\sigma$: $G(t,t_p)=\exp[-(t-t_p)^2/(2\sigma^2)]/(\sigma\sqrt{2\pi})$.
The absorption probability is proportional to the electron occupation and given by 
\begin{eqnarray}
Q(\varepsilon,\Omega_{\rm in})
  &=&
- f(\varepsilon) [1- f(\varepsilon + \hbar\Omega_{\rm in})] 
N(\varepsilon + \hbar\Omega_{\rm in})
\nonumber\\
 &+& [1-f(\varepsilon)] f(\varepsilon - \hbar\Omega_{\rm in})] 
 N(\varepsilon - \hbar\Omega_{\rm in}),
 \label{eq:laser2}
\end{eqnarray}
where $\hbar\Omega_{\rm in}$ is the photon energy. The factors $N(\varepsilon - \hbar\Omega_{\rm in})$ and $N(\varepsilon + \hbar\Omega_{\rm in})$ are needed to satisfy the electron number conservation during the electron excitation. The use of Eqs.~(\ref{eq:laser1}) and (\ref{eq:laser2}) enables to observe a cascade-type excitation in the electron distribution: The absorption of multiple photons occurs in a step-by-step manner, creating a high energy electron with, for example, $\varepsilon = \varepsilon_{\rm F} + m\hbar\Omega_{\rm in} \ (m\ge 2)$ (see Fig.~\ref{fig2} below). Similar nonequilibrium distribution function has been reported in numerical simulations \cite{mueller} and experiments \cite{fann}. 

%%%%%%%%%%%%%%%%%%%%%%%%%%%%%%%%%%%%
%\subsection{{\color{red}Photoluminescence intensity (old omega^3)}}
%The PL intensity at an energy $\hbar\Omega_{\rm out}$ is defined as \cite{suemoto2019}
%\begin{eqnarray}
%L(\Omega_{\rm out},t) &=& c_0 (\hbar\Omega_{\rm out})^3 \int d\varepsilon
%f(\varepsilon) [1-f(\varepsilon - \hbar\Omega_{\rm out})]
%\nonumber\\
%&\times&
%N(\varepsilon) N(\varepsilon - \hbar\Omega_{\rm out}),
%\label{eq:Lout_old}
%\end{eqnarray}
%where the factor of $(\hbar\Omega_{\rm out})^3$ comes from the Einstein relation. The coefficient $c_0$ depends on the material as well as the surface morphology but can be treated as a constant independent of $\hbar\Omega_{\rm out}$ in the first approximation. To model the experimental situation, the PL signal should be convoluted by a Gaussian pulse again
%\begin{eqnarray}
%L_{\rm conv}(\Omega_{\rm out}, t) =
%\int dt' G(t,t') L(\Omega_{\rm out},t').
%\label{eq:Lconv_old}
%\end{eqnarray}

\subsection{Photoluminescence intensity}
\label{sec:PL}
The photon emission from a metal with nonequilibrium electron distribution would be an analog to black-body radiation in thermal equilibrium, while the well-known Planck formula cannot be applied to the former situation. Instead of deriving the photon emission rate for the nonequilibrium situation from first-principles, we here propose a formula for the PL intensity that can be reduced to the Planck formula, the product of the photon DOS in vacuum and the expectation value of the photon number at $T=T_{\rm e}$, in the limit of electron quasiequilibrium. We define the PL intensity at an energy $\hbar\Omega_{\rm out}$ as
\begin{eqnarray}
L(\Omega_{\rm out},t) &=& c_0 (\hbar\Omega_{\rm out})^2 J(\Omega_{\rm out}),
\label{eq:Lout}
\\ 
J(\Omega_{\rm out}) &=&
\frac{1}{\hbar\Omega_{\rm out}}
\int_{-\infty}^{\infty} d\varepsilon
f(\varepsilon) [1-f(\varepsilon - \hbar\Omega_{\rm out})]
\nonumber\\
&\times&
{\tilde N}(\varepsilon) {\tilde N}(\varepsilon - \hbar\Omega_{\rm out}),
\label{eq:Jout}
\end{eqnarray}
where the factor of $(\Omega_{\rm out})^2$ in Eq.~(\ref{eq:Lout}) corresponds to the photon DOS. ${\tilde N}(\varepsilon) = N(\varepsilon)/N(\varepsilon_{\rm F})$ is the normalized electron DOS. Assuming that $f(\varepsilon)$ is equal to the Fermi-Dirac (FD) distribution with $T_{\rm e}$ and that $N(\varepsilon)$ is constant around $\varepsilon_{\rm F}$, where the latter is a good approximation for the bandstructure in noble metals, $J(\Omega_{\rm out})$ can be equal to the Bose-Einstein (BE) function:
\begin{eqnarray}
%\int d\varepsilon f(\varepsilon) [1-f(\varepsilon - \hbar\Omega_{\rm out})]
J(\Omega_{\rm out})
= \left[ \exp
\left( \frac{\hbar\Omega_{\rm out}}{k_{\rm B}T_{\rm e}} \right) - 1
\right]^{-1},
\end{eqnarray}
where $k_{\rm B}$ is the Boltzmann constant. Thus Eq.~(\ref{eq:Lout}) is reduced to the Planck formula for the photon number. Given these assumption, $L(\Omega_{\rm out},t)$ takes a maximum value at the photon energy of $\hbar\Omega_{\rm out}\simeq 1.59 k_{\rm B}T_{\rm e}$, known as the Wien's displacement law for the photon number. A similar discussion for deriving the PL intensity of metals has been given in Ref.~\cite{haug}. In this way, the black-body spectra have been derived exactly. For possible future use, we added the coefficient $c_0$ in Eq.~(\ref{eq:Lout}). It corrects the discrepancy of PL spectra between the ideal and realistic samples and may have the energy dependence, although we have no theory for determining $c_0$ accurately. In the present simulation we treated it as a constant independent of $\hbar\Omega_{\rm out}$.

To model the experimental situation, the PL signal should be convoluted by a Gaussian pulse again
\begin{eqnarray}
L_{\rm conv}(\Omega_{\rm out}, t) =
\int dt' G(t,t') L(\Omega_{\rm out},t').
\label{eq:Lconv}
\end{eqnarray}

%\sout{Given that our sample can be considered as a black-body, the expression for the PL intensity is simply given by $L(\Omega_{\rm out},t) = (\hbar\Omega_{\rm out})^2 J(\Omega_{\rm out})$; Otherwise, some correction for the energy dependence should be added. The coefficient $c_0$ in Eq.~(\ref{eq:Lout}) serves as such a correction factor, although we have no theory for determining $c_0$ accurately.}  

%%%%%%%%%%%%%%%%%%%%%%%%%%%%%%%%%%%%
\subsection{Computational details}
\label{sec:comp}
%%%%%%%%%%%%%%%%%
\begin{table}[bbb]
\begin{center}
\caption{Numerical parameters for laser excitations}
{
\begin{tabular}{lc}\hline
%--------------------------------------------------------------------------------
 pulse intensity  \hspace{5mm}& $P_0$ (or $W$) \\ \hline
 pulse width  \hspace{5mm}& $\sigma=55$ fs (FWHM=$130$ fs) \\ \hline
 photon energy \hspace{5mm} & $\hbar\Omega_{\rm in}=1.19$ eV \\ \hline
% relaxation time (e-e collision) \hspace{3mm} & $\tau_{\rm e-e}$  \\ \hline
%--------------------------------------------------------------------------------
\end{tabular}
}
\label{tab:para}
\end{center}
\end{table}

We use density-functional theory and density-functional perturbation theory implemented into Quantum ESPRESSO code \cite{qe} to obtain $N(\varepsilon)$, $D(\omega)$ and $\alpha^2 F(\omega)$. The effects of exchange and correlation are treated within PBE-GGA \cite{pbe}. The core electrons are treated within the ultrasoft pseudopotential method \cite{uspp}. The cutoff energies for the wavefunction and the charge density are 60 Ry and 400 Ry, respectively. 
$N(\varepsilon)$ is obtained from a self-consistent (scf) calculation using 20$\times$20$\times$20 grids. $D(\omega)$ and $\alpha^2 F(\omega)$ are obtained from scf calculations using 30$\times$30$\times$30 $k$ grids including $k$ and $k+q$ points, 10$\times$10$\times$10 $k$ grids for phonon calculations, and 5$\times$5$\times$5 $q$ grids. 

The numerical parameters used for the laser-excitation term of Eq.~(\ref{eq:Boltzmann_el}) are listed in Table~\ref{tab:para}. Numerical integration of Eqs.~(\ref{eq:Boltzmann_el}) and (\ref{eq:Boltzmann_ph}) was performed by the time-step of $0.24$ fs and started at $t\ll t_p$ prior to laser irradiation. The electron energy window is $\varepsilon \in [\varepsilon_{\rm F}-2.5 \ {\rm eV},\varepsilon_{\rm F} + 2.5 \ {\rm eV}]$ that is discretized with $10^4$ grids. Since the $d$ band in silver is located below the Fermi level by about 3 eV, no $d$ electron excitations exist. The maximum phonon energy (Debye energy) of silver is 25 meV discretized with 50 grids. The energy interval is thus 0.5 meV for both the electron and phonon energies. With these numerical setting, we obtain the second moment of the Eliashberg function of $\lambda \langle \omega^2 \rangle=41$ meV$^2$, agreement with an estimation in Ref.~\cite{lin}. To reduce the computational cost of the e-e collision integral, Eq.~(\ref{eq:ee2}), we perform the integration at coarse grids of $10^3$ points and interpolate to the original $10^4$ points. 

%$\lambda = 0.13$ and 

The value of the first parameter $v_0$ in Eq.~(\ref{eq:ee2}) is inferred as follows: We assume that the e-e interaction potential is described by the Yukawa-type potential, whose Fourier-transform is given by
\begin{eqnarray}
 v_{\rm sc}(k) = \frac{1}{\Omega_{\rm cell}} \frac{e^2}{\varepsilon_{0} (k^2 + k_{\rm TF}^{2})},
 \label{eq:yukawa}
\end{eqnarray}
where $\Omega_{\rm cell}$ is the volume of the unit cell, $e$ is the elementary charge, $\varepsilon_0$ is the dielectric constant of vacuum, and $k_{\rm TF}$ is the Thomas-Fermi wavelength. We set $\Omega_{\rm cell}=16.96$ \AA$^3$ and $k_{\rm TF}=1.7$ \AA$^{-1}$ \cite{ashcroft}. Next, we assume that the wavenumber transfer $k$ occurs on average $2k_{\rm F}$ (i.e., the diameter of the Fermi sphere with a radius $k_{\rm F}$) in the e-e scattering events. By substituting $k=2k_{\rm F}$  into Eq.~(\ref{eq:yukawa}), one obtains $v_0 = 1.23$ eV. 

We define the excess electron and phonon energies as  
\begin{eqnarray}
E_{\rm e}
&=& 
2\int d\varepsilon \varepsilon N(\varepsilon) [f(\varepsilon) - f_{\rm FD}(\varepsilon,T_0)],
\label{eq:totE_e}
\\
E_{\rm ph}
&=& 
\int d\omega \hbar\omega D(\omega) [n(\omega) - n_{\rm BE}(\omega,T_0)],
\label{eq:totE_ph}
\end{eqnarray}
where $f_{\rm FD}(\varepsilon,T_0)$ and $n_{\rm BE}(\omega,T_0)$ are the FD and BE function at $k_{\rm B}T_0=0.025$ eV. The factor of $2$ in Eq.~(\ref{eq:totE_e}) comes from the spin degeneracy. When $t\gg t_p$, the total excess energy, $E_{\rm e} + E_{\rm ph}$, does not change with time due to the energy conservation. We define the excitation density as $W=E_{\rm e} + E_{\rm ph}$ in the limit of $t\rightarrow \infty$. The magnitude of $W$ increases when $P_0$ is increased and will be optimized to reproduce the experimental data. This is another parameter in the present study. 

Below we study the time-evolution of the PL spectra, $L_{\rm conv}(\Omega_{\rm out}, t)$ given by Eq.~(\ref{eq:Lconv}), in silver. In addition, we will focus on the transient PL at $\hbar\Omega_{\rm out}=0.9$ eV that is measured experimentally. Since the gaussian pump shows a peak at $t=t_p$ and a width $\sigma$, the PL intensity takes a maximum at around $t=t_0\simeq t_p + 2\sigma$ ps: The electrons are still excited by the photon absorption even after $t\ge t_p$ due to the finite $\sigma$. In the following, the time $t$ is measured from $t_0$ deduced from the PL decay at $\hbar\Omega_{\rm out}=0.9$ eV, while such $t_0$ is almost insensitive to $\hbar\Omega_{\rm out}$ in our numerical simulation. 

%%%%%%%%%%%%%%%%%
\begin{figure}[ttt]
\center
\includegraphics[scale=0.42]{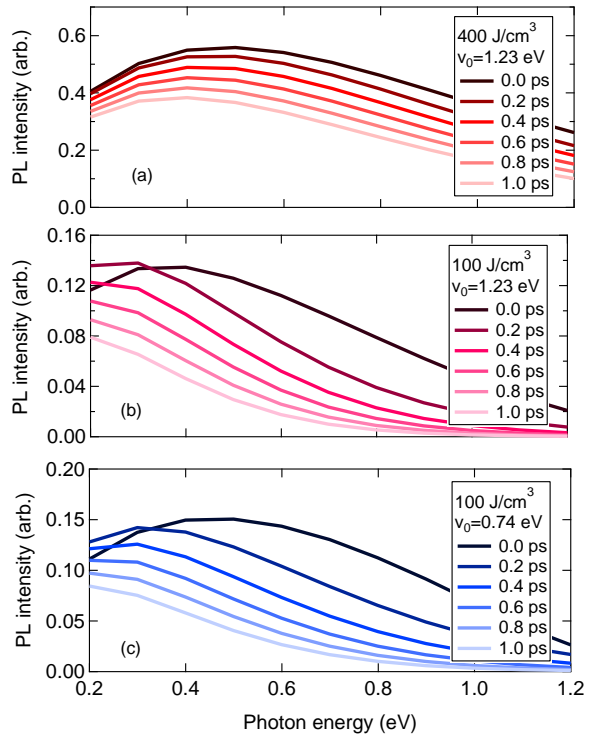}
\caption{\label{fig1} The time-evolution of the PL spectra in silver: (a) $W=400$ J/cm$^3$, $v_0=1.23$ eV, (b) $W=100$ J/cm$^3$, $v_0=1.23$ eV, and (c) $W=100$ J/cm$^3$, $v_0=0.74$ eV. }
\end{figure}
%%%%%%%%%%%%%%%%%
%%%%%%%%%%%%%%%%%
\begin{figure}[ttt]
\center
\includegraphics[scale=0.45]{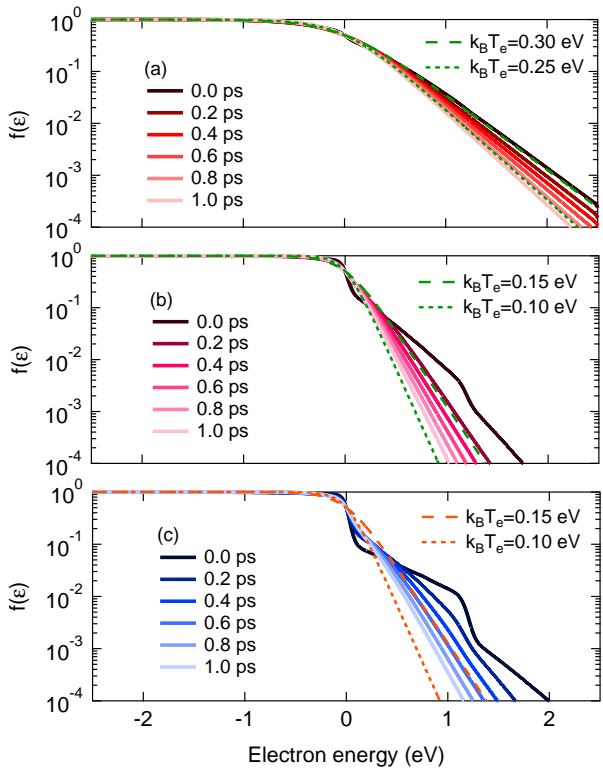}
\caption{\label{fig2} The time-evolution of $f(\varepsilon)$ for (a) $W=400$ J/cm$^3$, $v_0=1.23$ eV, (b) $W=100$ J/cm$^3$, $v_0=1.23$ eV, and (c) $W=100$ J/cm$^3$, $v_0=0.74$ eV. The electron energy is measured from $\varepsilon_{\rm F}$. The dashed and dotted curves are the FD function with $T_{\rm e}$, whose values are estimated from the total electron energy at $t=0.0$ ps and 1.0 ps.}
\end{figure}
%%%%%%%%%%%%%%%%%

%%%%%%%%%%%%%%%%%%%%%%%%%%%%%%%%%%%%
\section{Results and Discussion}
%%%%%%%%%%%%%%%%%%%%%%%%%%%%%%%%%%%%
%%%%%%%%%%%%%%%%%%%%%%%%%%%%%%%%%%%%
\subsection{Time-evolution of PL spectra}
\label{sec:R1}
%%%%%%%%%%%%%%%%%%%%%%%%%%%%%%%%%%%%

Figures \ref{fig1}(a) and \ref{fig1}(b) show the time-evolution of $L_{\rm conv}(t)$ for excitation densities; (a) $W=400$ J/cm$^3$ and (b) $W=100$ J/cm$^3$. The value of $v_0$ is set to be 1.23 eV. For $W=400$ J/cm$^3$, the curve of $L_{\rm conv}(t=0)$ takes a maximum at around $\hbar\Omega_{\rm out}\simeq0.5$ eV. The peak shifts to lower energies with time because the excited electrons relax toward the Fermi level by transferring their energy to phonons. For $W=100$ J/cm$^3$, the peak of $L_{\rm conv}(t=0)$ is located at $\hbar\Omega_{\rm out}=0.4$ eV that is shifted toward lower energies again. Note that the photon energy at which $L_{\rm conv}(t=0)$ takes a maximum value is smaller than the case of $W=400$ J/cm$^3$. This is because the magnitude of high energy tail of $f(\varepsilon)$ becomes small in response to the decreased averaged electron energy. Figures \ref{fig2}(a) and \ref{fig2}(b) show the time-evolution of $f(\varepsilon)$ for $W=400$ J/cm$^3$ and $W=100$ J/cm$^3$, respectively. The electron quasiequilibrium is almost kept for both $W$s: $f(\varepsilon)$ is well fitted by $f_{\rm FD}(\varepsilon,T_{\rm e})$ with $k_{\rm B}T_{\rm e} \in [0.25,0.30]$ eV within an interval of $t\in [0.0, 1.0]$ ps and with $k_{\rm B}T_{\rm e} \in [0.10,0.15]$ eV within $t\in [0.2, 1.0]$ ps for cases (a) and (b), respectively, where the value of $T_{\rm e}$ is estimated from the total electron energy. This is because the e-e scattering rate proportional to $v_{0}^{2}$ is strong enough to allow the nonequilibrium component created by photon absorption to be smeared out at any time, while non-negligible deviation from $f_{\rm FD}(\varepsilon,T_{\rm e})$ with $k_{\rm B}T_{\rm e}=0.15$ eV is observed at $t=0$ ps for the case (b), shown in Fig.~\ref{fig2}(b). What is important here is that the decrease in the high energy tail of $f(\varepsilon)$ can cause the redshift in the PL spectra. 

To understand how nonequilibrium electron distribution influences the PL spectra, we set $v_0=0.74$ eV by assuming $k = 2k_{\rm TF}$ in Eq.~(\ref{eq:yukawa}) and show the time-evolution of $L_{\rm conv}(t)$ for $W=100$ J/cm$^3$ in Fig.~\ref{fig1}(c). Compared to Fig.~\ref{fig1}(b), the peaks of $L_{\rm conv}(t)$ are shifted to larger energies. Since the e-e scattering rate is suppressed compared to the case of $v_0=1.23$ eV, the PL peak around $\hbar\Omega_{\rm out}=0.5$ eV at $t=0$ ps must be attributed to the nonequilibrium electrons created by the laser absorption. In fact, the stepwise electron distribution is more clearly observed when $t\simeq 0$ ps, shown in Fig.~\ref{fig2}(c). The high energy electrons created by cascade-type excitations can contribute to the blueshift of the PL peak. These comparative studies indicate that the time-evolution of PL spectra reflects the relaxation dynamics of electron distribution function. 

%%%%%%%%%%%%%%%%%%%%%%%%%%%%%%%%%%%%
\subsection{Application to experiment}
\label{sec:R2}
%%%%%%%%%%%%%%%%%%%%%%%%%%%%%%%%%%%%

%%%%%%%%%%%%%%%%%
\begin{figure}[ttt]
\center
\includegraphics[scale=0.45]{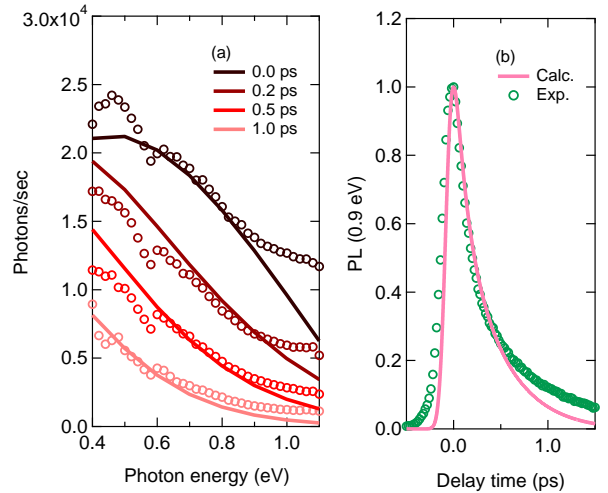}
\caption{\label{fig3} (a) The time-evolution of the PL spectra of silver: The experimental PL (circle) and the calculated PL (solid) assuming $W=100$ J/cm$^3$, $v_0=0.74$ eV. (b) Comparison of the transient PL at $\hbar\Omega_{\rm out}=0.9$ eV.}
\end{figure}
%%%%%%%%%%%%%%%%%
%%%%%%%%%%%%%%%%%
\begin{figure}[ttt]
\center
\includegraphics[scale=0.4]{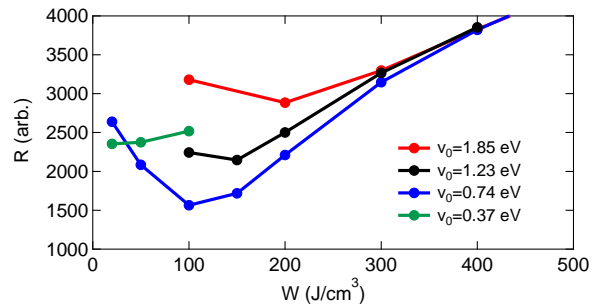}
\caption{\label{fig4} $R$ versus $W$ for $v_0=1.85, 1.23, 0.74$, and 0.37 eV.}
\end{figure}
%%%%%%%%%%%%%%%%%

We apply our theory to experiment. Figure \ref{fig3}(a) shows the experimental data of PL spectra in silver (circle). Details of the experimental setup are given elsewhere \cite{suemoto2019}, while a correction of the energy dependence of the emissivity has been employed in the present work. The PL spectra takes the maximum value at around 0.5 eV when $t=0$ ps. The PL intensity decreases rapidly with time: Within 1 ps, the PL intensity becomes less than about one fifth.

%{\color{red}\sout{The peak position of the calculated PL is larger than that of the experiment. In addition, }} it is clear that the present model never explains the experimental data: 

%(i) an overestimation of the excitation density $W$ and, more importantly, (ii)
% As shown in Figs.~\ref{fig1}(a) and \ref{fig1}(b), the smaller $W$, the lower the photon energy at which the curve of $L_{\rm conv}(t=0)$ takes a maximum. To reproduce the experimental observation, we thus set $W=200$ J/cm$^3$ in the present study. Nevertheless, the PL decay is still much slower than the experiment. The discrepancy between the theory and experiment in silver may be due to a lack of the effect of the excess energy diffusion. 

We compare the calculated PL spectra with the experiment quantitatively as follows. First, we scale the magnitude of the calculated PL to minimize the standard deviation \cite{minpack}
\begin{eqnarray}
 R^{2} = \frac{1}{N_{\rm data}} \sum_{j,k}
  \left[ L_{\rm conv}^{\rm (calc)}(\Omega_{j},t_k) 
         - L_{\rm conv}^{\rm (exp)}(\Omega_{j},t_k) \right]^2,
         \label{eq:SD}
\end{eqnarray}
where the summations with respect to $j$ and $k$ extend over the PL energy ($\hbar\Omega = 0.4$ to 1.1 eV with an increment of 0.1 eV) and the delay time ($t=0.0, 0.2, 0.5, 1.0$ ps). $N_{\rm data}$ is the number of data points used in the fitting, i.e., $N_{\rm data} = 32$. Next, we adjust two parameters, $v_0$ and $W$. We choose $v_0=1.85, 1.23, 0.74$, and $0.37$ eV that correspond to $k=k_{\rm TF}, 2k_{\rm F}, 2k_{\rm TF}$, and $3k_{\rm TF}$, respectively, in Eq.~(\ref{eq:yukawa}). For a rough estimate of $W$, we assumed that the absorption length of the laser light is about 10 nm. The pulse fluence is estimated to be 0.58 mJ/cm$^2$ from the laser spot size (18 $\mu$m) on the sample surface and the absorption rate (80 \%). The value of $W$ is thus estimated to be 460 J/cm$^3$. In addition, we study the several values of $W\in [20, 460]$ J/cm$^3$, depending on the value of $v_0$. Figure \ref{fig4} shows the estimated $R$ given by Eq.~(\ref{eq:SD}) as a function of $W$ for various $v_0$s. The optimized parameters are $v_0=0.74$ eV and $W=100$ J/cm$^3$, giving $R\simeq1500$. 

The PL spectra using the optimized parameters are shown in Fig.~\ref{fig3}(a) (solid curve). The agreement is good except for the photon energy larger than 1.0 eV. Figure \ref{fig3}(b) shows a comparison to the transient PL intensity at $\hbar\Omega_{\rm out}=0.9$ eV. It is clear that the calculated PL decay is in agreement with the decay observed in experiment. 

The present analysis implies that the contribution from nonequilibrium electron is important to understand the transient PL in silver. With the optimized parameters, the time-evolution of electron distribution is again given in Fig.~\ref{fig2}(c). As $v_0$ increases, the deviation from the FD function becomes small, as shown in Fig.~\ref{fig2}(b) for the case $1.23$ eV, which gives rise to an increase in $R$, as shown in Fig.~\ref{fig4}. When $v_0$ is decreased from the optimized value, the magnitude of $R$ also increases because of strong electron nonequilibrium.  

The optimized $W$ is about one-fourth of the initial guess (460 J/cm$^3$). This is validated by considering that there are nanoscale roughness in the sample surface. The surface area with such roughness becomes large effectively, compared to the laser spot size, yielding a small value of $W$. The optimized $v_0$ is derived from the wavenumber transfer of $2k_{\rm TF}\simeq 3.4$ \AA$^{-1}$ in the e-e scattering on average. This value is larger than $2k_{\rm F}\simeq 2.4$ \AA$^{-1}$ and the size of the reciprocal lattice vector $4\pi/a_{\rm lat} \simeq 3.1$ \AA$^{-1}$ \ with $a_{\rm lat}$ being the lattice constant. This implies that the contribution from e-e umklapp scattering is not negligible. To study the impact of umklapp scattering on the electron dynamics in detail, the Boltzmann equation must be solved by extending Eq.~(\ref{eq:cee}) to include the umklapp process and assuming not the energy-space but the wavevector-space grid for the distribution function.

Although the present model captures the overall features of the PL spectra, some deviations between the theory and experiment are present; In Fig.~\ref{fig3}(a), the theoretical curves at high energy underestimates the magnitude of PL; and in Fig.~\ref{fig3}(b), the PL decay is fast when the delay time is larger than 1.0 ps. One possible explanation is that the rough surface would contain traps that can capture high energy electrons. Such an electron localized state around the Fermi level will change the e-e and e-ph scattering rates from the bulk values that are calculated in the present work and therefore will slow the electron relaxation. The detailed study of the effect of inhomogeneity on the electron dynamics will be left for future work. In addition, a microscopic theory for PL of metals has to be developed by considering the surface plasmon resonance \cite{suemoto2019,beversluis}. 

\section{Conclusion}
\label{sec:conclusion}
%%%%%%%%%%%%%%%%%%%%%%%%%%%%%%%%%%%%
We have calculated ultrafast PL of silver by solving the Boltzmann equation taking into account the effect of e-e and e-ph collisions. By comparing with experiment, we have demonstrated that the situation of electron nonequilibrium is important in understanding the transient PL spectra in silver. To improve the agreement between the theory and experiment, more investigations are required to understand several effects such as nanoscale roughness at metal surfaces and the e-e umklapp scattering on the ultrafast electron dynamics. We expect that our work motivates the study of such dynamics in realistic systems as well as the PL decay in photoexcited metals.

%The calculated PL decay is much slower than that in experiment. 

%We have investigated the femtosecond PL dynamics of metals using first-principles calculations and solving the Boltzmann equation. The fast and slow decay components are observed in the PL dynamics. Through a systematic investigation of the effect of e-ph, ph-ph, and e-e scattering, we have shown that the slow decay component reflects the presence of nonthermal electrons. 

\begin{acknowledgments}
One of the authors (T.S.) would like to thank Dr. K. Yamanaka and Mr. N. Sugimoto at Toyota Central R\&D Labs., Inc. for the collaboration with spectroscopic measurements and surface characterization. This study is supported by the Nikki-Saneyoshi Foundation and a Grant-in-Aid for Scientific Research (C) (Grant No. 17K05505) from JSPS. A part of numerical calculations has been done using the facilities of the Supercomputer Center, the Institute for Solid State Physics, the University of Tokyo.
\end{acknowledgments}

%\appendix
%\section*{Appendix}

\end{document}